# Superconductivity-driven magnetization modulation in $YBa_2Cu_3O_{7-\delta}$ /$SrTiO_3$/$La_{0.67}Sr_{0.33}MnO_3$ heterostructures

Surendra Singh[1,2,*], Harsh Bhatt[1], Yogesh Kumar[1], C. L. Prajapat[3], B. Satpati[4], C. J. Kinane[5], S. Langridge[5], G. Ravikumar[2,6] and S. Basu[1,2]

[1]Solid State Physics Division, Bhabha Atomic Research Centre, Mumbai 400085 India

[2]Homi Bhabha National Institute, Anushaktinager, Mumbai 400094 India

[3]Technical Physics Division, Bhabha Atomic Research Centre, Mumbai 400085 India

[4]Surface Physics and Material Science Division, Saha Institute of Nuclear Physics, 1/AF, Bidhannagar, Kolkata 700064, India

[5]ISIS-Neutron and Muon Source, Rutherford Appleton Laboratory, Didcot, Oxon OX11 0QX, United Kingdom

[6] Scientific Information Resource Division, Bhabha Atomic Research Centre, Mumbai 400085 India

*surendra@barc.gov.in

**Abstract:** Using spin polarized neutron reflectivity experiments, we demonstrate an unusual proximity behaviour when the superconductor (SC) and the ferromagnet (FM) are coupled through an insulator (I) in $YBa_2Cu_3O_{7-\delta}$ (SC)/$SrTiO_3$ (I)/$La_{0.67}Sr_{0.33}MnO_3$ (FM) heterostructures. We have observed an unexpected magnetic modulation at the interface region of the FM below the superconducting transition temperature. The magnetization of the FM layer at the I/FM interface was drastically reduced as compared to the magnetization in the rest of the FM layer. This result indicates that the Cooper pairs tunnel across the insulator and interact with the local magnetization at the interface region (extending ~ 30 Å) of the FM causing modification of the magnetization at the interface. This unexpected magnetic behavior cannot be explained on the basis of the existing theoretical models. However, the length scale associated here clearly suggests the long range proximity effect as a result of tunneling of Cooper pairs.

**Introduction:**

The proximity effect in superconductor (SC)/ferromagnet (FM) hybrid systems has attracted considerable interest as a result of the fascinating basic physics and the possibility to exploit such phenomena for promising applications in superconducting spintronics [1-6]. Earlier studies on SC/FM hybrid systems have combined superconductivity and magnetism and have mainly focused on the injection of spin-polarized quasiparticles into the SC. Subsequently, the tunneling/injection of spin polarized quasiparticles from a FM to a SC through an insulator (I) in SC/I/FM hybrid oxide heterostructures have been explored with a view to combining spintronics and superconductivity, into the so-called super-spintronics [7-12]. However, understanding the phenomenon of superconducting carriers tunneling into the FM and its effect on the magnetic state of the FM layer in SC/I/FM oxide heterostructure has largely remained elusive. It is well understood, for s-wave and d-wave superconductors in SC/FM systems, that the spin-singlet Cooper pairs (S = 0) have a limited coherence length even for a weak FM [1], owing to the magnetic exchange interaction. The Josephson coupling between SCs separated by a thick FM layer clearly suggests the possibility of superconducting order penetrating into strong FMs [13-17]. The long-range superconducting order at FM/SC interface was attributed to the presence of spin- triplet Cooper pairs (S = 1) at the interfaces, which are not as sensitive to the exchange field and largely depend on the existence of magnetic inhomogeneities, such as ferromagnetic domain walls or noncollinear magnetization at the interfaces [2-4].

The most studied SC/FM oxide interfaces are $YBa_2Cu_3O_{7-\delta}$ (YBCO), a d-wave SC, and half-metallic manganites, like $La_{0.67}Ca_{0.33}MnO_3$ (LCMO) or $La_{0.67}Sr_{0.33}MnO_3$ (LSMO), as the FM. Several intriguing interfacial phenomena have been reported for these SC/FM heterostructures [18-26]. However, there are mixed reports regarding the appearance of spin-triplet Cooper pairs in the YBCO/LCMO (or LSMO) systems [27-29]. Recently, Visani *et al.*,[30] demonstrated long-range proximity effects between YBCO and LCMO due to the interference effects between a quasiparticle and electrons in the conductance spectra across the YBCO/LCMO interface and the correlation length was found to be as high as the thickness of the LCMO layer.

The key factor that has been established behind the observed long-range propagation of superconducting correlations into an half metal based FM/SC system is a conversion from spin-singlet to spin-triplet pairing near the interfaces [31, 32]. Introducing an insulator layer between the FM and SC is expected to hinder the propagation of long range superconducting correlations into the FM, unless there is tunneling of Cooper pairs through the insulator. Using polarized neutron reflectivity (PNR), we recently found strong magnetic modulation below the superconducting transition temperature ($T_{SC}$) in YBCO/$SrTiO_3$(STO)/LCMO hybrid heterostructures grown on STO substrates [33]. While PNR results clearly indicated a magnetic modification at the interfaces, there was uncertainty involved because of the closeness of the magnetic transition temperature ($T_c$) of LCMO with the structural phase transition of STO [34, 35] in this system. Recently Paull et. al. [36] also confirmed the magnetic modulation in the YBCO/STO/LCMO heterostructure below $T_{SC}$.

Here we report a drastic reduction in magnetization at the interface of a LSMO layer through an insulator (STO) in YBCO/STO/LSMO heterostructures grown on MgO single crystals, using depth-sensitive PNR technique [33, 36-40], below $T_{SC}$. PNR showed a strong reduction in interface magnetization with a small negative magnetization (magnetization opposite to the applied field) in the LSMO layer over a thickness of ~ 30±5 Å, at the STO/LSMO interface in YBCO/STO/LSMO heterostructure below $T_{SC}$. A suppression of the magnetization at STO/LSMO interface was also observed at 100 and 200 K (well above $T_{SC}$). PNR measurements, from similarly grown STO/LSMO bilayer also showed a reduction in the interfacial magnetization at 100 and 200 K, suggesting that the reduced magnetization above $T_{SC}$ at the STO/LSMO interface may be an intrinsic phenomenon. However, depleted magnetic interface with small negative magnetization at STO/LSMO interface below $T_{SC}$ in YBCO/STO/LSMO heterostructures is solely dependent on superconductivity.

**Experimental:**

We grew a number of hybrid heterostructures by pulsed laser deposition on single crystal MgO substrates. In Table 1, we provide a list of heterostructures studied here, two trilayers (YSL25 & YSL50) and two bilayers (SL & YL). The letters Y, S and L stand for YBCO, STO and LSMO respectively. Two (three) alphabets in a sample code indicates a bilayer (trilayer). The numbers 25 and 50 in the trilayer codes are the nominal thicknesses of the insulating STO layer measured in Å. For example, YSL25 identifies a sample with deposition sequence of YBCO followed by 25 Å STO layer and then LSMO and the heterostructure can be represented as: MgO/YBCO/STO(25 Å)/LSMO. An excimer laser (KrF, λ=248 nm, pulse width = 20 nm) was used to ablate high density targets of LSMO, YBCO and STO. The laser fluence was fixed at 3 J/cm$^2$ and an optimized pulse repetition rate of 2 Hz was used for the deposition under an oxygen partial pressure of 0.5 mbar. The film deposition was carried out at an optimized substrate temperature of 800 °C.

Magnetization measurements were performed using Quantum Design superconducting quantum interference device (SQUID) magnetometer MPMS5 under field cooled (FC) and zero field cooled (ZFC) conditions. The morphology and structural properties of the layers were separately investigated by scanning TEM and high-resolution TEM. The cross-sectional TEM specimens were prepared using conventional method by mechanical grinding followed by dimpling (down to below 20 μm) and low-energy (3 keV) and low-angle (4°) Ar-ion milling. Final cleaning was done by 1.2 keV Ar ion milling. TEM images were acquired using a FEI, Tecnai G2 F30, S-Twin microscope operating at 300 kV equipped with a Gatan Orius CCD camera. X-ray diffraction (XRD) and X-ray reflectivity (XRR) were carried out using Cu K$_\alpha$ radiation in a rotating anode source. PNR experiments were performed on the POLREF reflectometer at the ISIS facility, RAL, UK.

The specular (angle of incidence = angle of reflection) reflectivity ($R$) was measured as a function of wave vector transfer, $Q = 4\pi \sin\theta/\lambda$ (where, $\theta$ is angle of incidence and $\lambda$ is the wavelength of the probe, i. e. x-ray/neutron). The reflectivity is qualitatively related to the Fourier transform of the scattering length density (SLD) depth profile $\rho(z)$ [37-40], averaged over whole sample area. For XRR, $\rho(z)$, is proportional to electron density[37-39]. For PNR, $\rho(z)$ consists of nuclear SLD (NSLD) and magnetic SLD (MSLD) such that $\rho^{\pm}(z) = \rho_n(z) \pm \rho_m(z) = \rho_n(z) \pm CM(z)$, where $C$ = 2.91×10$^{-9}$ Å$^{-2}$ (emu/cc)$^{-1}$ and $M(z)$ is the magnetization (a moment density obtained in emu/cc) depth profile[39]. The +(-) sign denotes neutron beam polarization parallel (opposite) to the applied field and corresponds to respective reflectivities, $R^{\pm}(Q)$. Thus, by measuring $R^+(Q)$ and $R^-(Q)$, $\rho_n(z)$ and $M(z)$ can be obtained separately. The reflectivities were calculated using the dynamical formalism of Parratt [41], and parameters of the model were adjusted for the goodness of fit [42]. Errors reported for parameters obtained from XRR measurements represent the perturbation of a parameter that increased goodness of fit corresponds to a 2σ error (95% confidence) [43]. The PNR data reported in this paper were taken in an applied in-plane magnetic field ($H_{ext}$) of 500 Oe at different temperatures upon warming the samples from the lowest temperature ~ 7 K, after the samples were cooled at the same field (~500 Oe) from 300 to 7 K.

**Results and Discussion:**

The x-ray diffraction (XRD) scans in log scale along the growth direction for YSL25, YSL50, YL and SL heterostructures are shown in Fig. 1(a). A comparison of XRD data from these heterostructures (Fig. 1(a)) suggest ordered and strongly textured growth along (00$l$) directions. Textured growths for these oxide heterostructures were also observed earlier while grown on STO substrates [33, 44]. The YBCO (002) diffraction peak shown in the inset to Fig. 1 (a) (not shown in the main figure) clearly shows the interference fringes (indicated by vertical arrow in the inset of Fig 1(a)) around the diffraction peak, clearly suggest the high quality of interfaces. The microstructure of the identically grown YBCO/STO/LSMO heterostructure is visible in the cross-sectional high-resolution transmission electron microscopy (TEM) images in Fig. 1(b), showing that a well defined atomically sharp interface formed at the boundary. Fig. 1(c) and (d) show selected area electron diffraction pattern covering the films of YBCO and LSMO, respectively. The observed sharp electron diffraction spots indicate cubic and good quality single crystal structure with the orientation relationship with the substrate. Fig. 1(e) shows depth profile of intensity of K-edge characteristics of different elements, suggesting well define hybrid heterostructure. A combination of XRD, TEM and selected-area electron diffraction from the YBCO and LSMO layers together with a depth dependent elemental concentration profile of each element, further confirms the formation of highly ordered heterostructures.

The depth dependent structure of these heterostructures has been studied using x-ray reflectivity (XRR) [37-40] measurements. Fig. 2(a) shows the XRR data and corresponding fits (solid lines) from the YSL25, YSL50, YL and SL heterostructures and the corresponding electron scattering length density (ESLD) depth profile extracted from the XRR data are shown in Fig. 2(b-e). The structural parameters (thickness, ESLD and roughness) obtained from XRR for different heterostructures are given in Table 2.

Fig. 2(f) shows the zero-field-cooled (ZFC) and field-cooled (FC) (cooling field $H_{FC}$ = 100 Oe) magnetization of YSL25 and YL heterostructures measured in a field $H_{FC}$ = 100 Oe. A typical hysteresis loops at 100 K from these heterostructures (YSL25 and YL) are shown in Fig. 2(g) and (h). The FC data indicates that the LSMO layer has a Curie temperature just below 300 K ($T_C \approx$ 290 K). The ZFC data shows $T_{SC} \approx$ 63 K for YSL25 and YL heterostructures, which is lower than the usual value of $T_{SC}$ of bulk YBCO ($\approx$ 90 K).

To explore the depth profile of the magnetization across $T_{SC}$, we have used the PNR technique, because of its depth sensitivity for ordered magnetism [37-40]. A schematic of the PNR geometry is shown in the inset of Fig. 3(a). The difference between $R^+$ and $R^-$ reflectivities contains information on the magnetic depth profile [37-40]. Fig. 3(a) shows the PNR data and the fits for YSL25 at 300 K. PNR data at 300 K from YSL25 indicate no ferromagnetism, as the difference $R^+ - R^- \sim 0.0$, which is consistent with the SQUID measurements (Fig. 2(f)). The NSLD depth profile at 300 K is shown in Fig. 3(f) and the structural parameters obtained from PNR are also given in Table 2, along with parameters obtained from XRR. The depth dependent SLD profiles obtained from XRR and PNR at room temperature closely match each other, indicating robustness of the fits obtained. The room temperature NSLD depth profile was used as input to fit the low-temperature magnetic profiles in all cases. PNR data (scattered data) along with corresponding fits (solid lines) for YSL25 at different temperatures across $T_{SC}$ is depicted in Fig. 3(b-e). The MSLD depth profiles were obtained from the fits to the PNR data at low temperatures and are plotted in Fig. 3(g-j), as a function of the depth from the surface. Comparison of MSLD profile near STO/LSMO interface at 100 and 10 K has been shown in the Fig. 3(k). PNR data from YSL25 below $T_{SC}$ of YBCO i.e. 10 K and 50 K [Figs. 3(i-j)] indicate a region (thickness ~ 30 Å) of near zero magnetization that is marginally negative at the STO/LSMO interface.

PNR data for YSL25 at 100 and 200 K (above $T_{SC}$) also show a suppression of the magnetization in the LSMO layer up to ~50 Å from the STO/LSMO interface, with a MSLD of $4.4 \times 10^{-7}$ Å$^{-2}$ and $3.5 \times 10^{-7}$ Å$^{-2}$ at 100 and 200 K, respectively. There are varied reports suggesting both a reduction and enhancement in magnetization of the LSMO near the STO/LSMO interface, which are attributed to strain, oxygen content, deposition conditions and charge discontinuity [45-47]. In order to account for these different magnetization behaviors of the LSMO layer on STO, we have considered several magnetization models to fit the PNR data for YSL25 at 100 K and shown in Fig. S3(a-d) of Supplemental Material [48]. Different magnetic models (Fig. S3(e-h) of Supplemental Material [48]) are statistically compared for the quality of each fits using a goodness-of-fit parameter ($\chi^2$) for the normalized spin asymmetry parameter [36]. We observed that the model of a reduced magnetization for the LSMO layer (thickness ~ 50 Å) at the STO/LSMO interface best describes the PNR data (we obtained the lowest $\chi^2$ ~ 1.88 corresponding to this model). The magnetization of the interfacial LSMO layer increases on decreasing the temperature whilst above $T_{SC}$. Remarkably, we obtained a drastically reduced magnetization with an unusual small negative MSLD ~ -1.0×10$^{-7}$ Å$^{-2}$ (~ -36±25 emu/cc) for a region of the LSMO layer of thickness ~30±5 Å at the STO/LSMO interface below $T_{SC}$ (at 50 and 10 K). A comparison of fits of the PNR data at 10 K for YSL25, assuming different magnetization models are shown in Fig. S4 of Supplemental Material [48]. We considered several magnetization models: positive but reduced, zero and small negative magnetization for the interfacial LSMO layer. A significantly better fit ($\chi^2$ =1.9) for a small negative magnetization model as compared to the zero ($\chi^2$ = 2.7) and positive ($\chi^2$ = 3.5) magnetization models was observed. However there is large error (shown as vertical red lines in Fig 3(i-k)) for negative magnetization model below $T_{SC}$. Keeping in view the error limits on the fitted parameters, a drastic reduction in interfacial magnetization below $T_{SC}$ is clearly indicated in the PNR data at 10 K and 50 K. A comparison of reflectivity data at 100 K and 10 K can be seen as the ratio of R$^+$(10 K)/R$^+$(100 K) and R$^-$(10 K)/R$^-$(100 K) plotted in Fig S5(a) and (b) of Supplemental Material [48], respectively, suggesting significant difference in spin dependent reflectivity. The emergence of unusual magnetization for interfacial LSMO layer below $T_{SC}$ is significantly different from rest of the LSMO layer (~380±25 emu/cc). This unexpected magnetic behavior cannot be explained on the basis of the existing theoretical models. However it clearly suggests a long range proximity effect in this FM/I/SC system with tunneling geometry, which is suppressing magnetic order near interface in the heterostructure below $T_{SC}$.

In order to see the effect of the insulating layer thickness on the magnetization depth profile of the YBCO/STO/LSMO heterostructures across $T_{SC}$, we also studied another trilayer YSL50, with a thicker (~ 50 Å) STO layer. Figs. 4(a-c) show the PNR data from YSL50 at 300, 100 and 10 K. Fig. 4 (d) shows the corresponding NSLD depth profile and Fig. 4(e) and (f) show the depth profile of MSLD at 100 and 10 K, respectively, obtained from PNR data for the YSL50. Similar to YSL25, we obtained a reduced magnetization at the STO/LSMO interface at a temperature (~ 100 K) above $T_{SC}$ and a marginally negative magnetization at temperatures (10 K) below $T_{SC}$. The thickness of the LSMO layer at STO/LSMO interface, which shows drastically reduced magnetization with a small negative value (error bar: a vertical red line), has reduced to ~22±5 Å, whereas the value of magnetization at interface remains the same (~ -36 emu/cc), as obtained for YSL25. The length scales (thickness of STO layer and interface LSMO layer with reduced magnetization) associated with these two systems, YSL25 and YSL50, imply that the SC correlation may involves a flow of Cooper pairs and is thus a direct consequence of having long-range proximity effect through tunneling.

To further understand and establish the conventional proximity effect at the YBCO/LSMO interface, we have studied the structure-magnetic correlation of a YBCO/LSMO bilayer on MgO substrate (YL heterostructure) across $T_{SC}$. This heterostructure has also been characterized for structure (XRD) and macroscopic magnetization studies and showed high quality crystalline structure (Fig. 2). PNR measurements for YL, under similar conditions as adopted for YSL25 and YSL50, were also carried out at different temperatures across $T_{SC}$ (Fig. 5(a-c)). NSLD and MSLD depth profile of the YL at different temperatures obtained from the fits to PNR data are shown in Fig. 5(d-f). There are two distinctive features we observed in YL with respect to the trilayers. Firstly, we observed a uniform magnetization (Fig. 5(b) and (e)) for LSMO layer above $T_{SC}$ (100 K) and secondly we obtained a small negative magnetization (~ -35 emu/cc) over a much larger thickness (~ 60±7 Å) of LSMO layer at the YBCO/LSMO interface (Fig. 5(c) and (f)) below $T_{SC}$ (10 K). Magnetization change over a length scale of ~ 60 Å below $T_{SC}$ in the YL (proximity geometry) further reinforces the possibility of long range Cooper pairs.

Having observed the reduction in magnetization of interfacial LSMO layer at STO/LSMO interface in YSL25 and YSL50, above $T_{SC}$, the temperature dependent magnetization depth profile of the STO/LSMO interface without a SC layer will be another important question to address here. In this direction we studied a bilayer STO/LSMO (SL) heterostructure. Structural characterization (XRR and XRD data in Fig. 2) of SL sample confirms a high quality textured heterostructure. Fig. 6(a -d) show the PNR data from SL at different temperatures. Fig. 6(e) shows the NSLD depth profile of the SL heterostructure. Fig 6(f -h) show the MSLD depth profiles at different temperatures. Temperature dependent PNR data clearly suggest that a reduced magnetization for interfacial LSMO layer of thickness ~ 48±6 Å, best describes the measurements. Different magnetization models, e.g. uniform magnetization throughout LSMO layer, enhanced and reduced magnetization at interfaces were also considered to fit the PNR data at 100 and 10 K and given in Fig. S6 in Supplemental Material [48]. Comparison of the fits for different models clearly suggested that reduced magnetization model best describes the PNR data at different temperatures.

It is noted that the magnetization is reduced at the LSMO/STO interface in SL at both 200 and 100 K, which implies that the reduction in magnetization is due to the intrinsic property of LSMO/STO interface and not just due to the phase transition of STO near 110 K [34, 35]. To correlate the reduced magnetization at STO/LSMO interface with strain, we estimated the strain for the LSMO layer grown in different heterostructures on MgO substrate (Table S2 in the Supplemental Material [40]), which did not show any correlation of strain with reduced magnetization. However, other effects [45-47] at interface might have contributed for the reduced magnetization in SL (at all temperatures) and YSL (above $T_{SC}$) heterostructures. PNR data from SL also suggested that the magnetization of the LSMO/STO interface in SL increases on decreasing the temperature. In contrast the magnetization at LSMO/STO interface in YSL heterostructures showed negative (reduced) magnetization below (above) $T_{SC}$, suggesting superconductivity dependent phenomena.

The SLD for different oxide layers, obtained by XRR and PNR, are within 94-98% of their bulk values. However we obtained a magnetic SLD of ~$1.0 \times 10^{-6}$ Å$^{-2}$ (~ 400 emu/cc) at 10 K for LSMO layers, which is in agreement with earlier measurements for 40 nm LSMO film [49] but lower than its bulk value (~550 emu/cc) [47]. We obtained lower $T_C$ and $T_{SC}$ for these systems than their bulk values [$T_C$ ($T_{SC}$) for bulk LSMO (YBCO) ~ 330 K (92 K)]. $T_C$ of LSMO and $T_{SC}$ of YBCO vary with growth conditions, thickness of the film and type of substrates etc. However lower value of $T_C$ for LSMO films in present study is consistent with the $T_C$ (270 K-320 K) observed in different LSMO films [38, 45-47]. Lower $T_{SC}$ of YBCO film grown on MgO substrate were also reported earlier [50] and consistent with other hybrid heterostructures [25, 33, 44]. We believe that a lower value of $T_{SC}$, $T_C$ and magnetic moment in these systems are important parameters contributing to the observed results, because these may well provide additional energies and length scales that must be considered in describing the competing SC and magnetic interactions [51].

Negative magnetization in the SC layer in SC-FM-SC system was earlier ascribed to the inverse proximity effect [2]. A recent study on YBCO/STO/LSMO multilayer grown on STO substrates, using X-ray magnetic circular dichroism, also suggested an induced magnetization on the Cu sites and a reduction in the magnetization of Mn [44]. Such a transfer of magnetic moment from FM to SC is very small (~ -0.04 emu/cc for an applied field of 300 Oe) [33], and below the detection limit of PNR. The inverse proximity effect for non oxide systems has been observed over a much wider length scale into the SC layer [52, 53]. Recently Mironov *et al.*[53] proposed a theory of long ranged electromagnetic proximity effect suggesting strong spread of magnetic field into SC from FM. Our results show the direct proximity effect for YBCO/LSMO system and direct proximity effect through tunneling in YBCO/STO/LSMO systems. However we observed a marginally negative magnetization for interfacial LSMO layer. We propose that Cooper pairs, quantum tunnel through the insulating STO layer (penetrate) from YBCO to LSMO in YBCO/STO/LSMO (YBCO/LSMO) system, and interact with the conduction band electrons (spin polarized) of the half-metallic LSMO layer near the interface. The interaction of Cooper pairs with the local magnetization (phase separation at different length scale) in LSMO layer, may cause a possibility of negative spin accumulation or a canting of spins near the interfaces (due to spin frustration arising from this interaction), giving rise to a negative magnetization at the interface layer in YBCO/STO/LSMO (YBCO/LSMO) heterostructures. The existence of spontaneous spin accumulation have also been reported in a Josephson junction between a spin-singlet and a spin-triplet superconductor [54] as well as in an SC/FM/SC Josephson junction with strong spin-orbit coupling in the FM layer [55]. The systematic study described here by characterizing different interfaces from a number of heterostructures and earlier studies on strongly textured oxide heterostructure YBCO/STO/LCMO grown on STO substrate [33, 36, 45] clearly generalize the observation of reduction of interfacial magnetization, which is driven by superconductivity in these oxide heterostructures. Conversion of magnetic depleted layer to ordered ferromagnetism for interfacial LCMO layer at low temperature in the YBCO/STO/LCMO heterostructure was earlier observed, however the high magnetic field applied to the heterostructure destroyed the superconducting properties of the system [36].

**Conclusion:**

In summary, we have experimentally observed drastically reduced with small negative magnetization for an interfacial LSMO layer in the YBCO/STO/LSMO trilayers (tunneling geometry) and in the YBCO/LSMO bilayer (proximity geometry) below $T_{SC}$. PNR data provided direct evidence that the YBCO/STO/LSMO and the YBCO/LSMO heterostructures exhibit SC driven negative interfacial magnetization for LSMO layer (thickness ~ 30-60 Å) at the STO/LSMO and the YBCO/LSMO interfaces, respectively. The length scale associated with this tunneling driven proximity effect strongly suggests the presence of long range Cooper pairs in this system. However future theoretical analysis is necessary to confirm this. Magnetization modulation in an SC/I/FM system driven by tunneling may serve a basis for envisaging device applications in superconducting spintronics.

**Acknowledgements**

Y. K., acknowledges the support from Department of Science and Technology (DST), India via the DST INSPIRE faculty research grant (DST/INSPIRE/04/2015/002938). We thank the ISIS Neutron and Muon source for the provision of beam time.

Table 1: Designed heterostructures with their magnetic and superconducting order temperatures.

| Sample Code | Designed structure of the Sample | Magnetic transition temperature, $T_c$ | Superconducting transition temperature, $T_{SC}$ |
|---|---|---|---|
| SL | MgO(substrate)/STO/LSMO | ~290 K | - |
| YSL25 | MgO(substrate)/YBCO/STO(25Å)/LSMO | | -~63 K |
| YL | MgO(substrate)/YBCO/LSMO | | |
| YSL50 | MgO(substrate)/YBCO/STO(50Å)/LSMO | | |

Table 2: Parameters obtained from XRR and PNR measurements for different heterostructures grown on MgO substrates. Parameters values given in parenthesis are from PNR.

| \multicolumn{4}{c}{**MgO(Sub.)/YBCO/STO(25 Å) /LSMO (YSL25) heterostructure**} |
|---|---|---|---|
| layer | Thickness (Å) | SLD ($10^{-5}$ Å$^{-2}$) | Roughness (Å) |
| LSMO | 200±8[207±8] | 4.67±0.06[0.37±0.02] | 12±3[10±3] |
| STO | 27±3[25±2] | 3.90±0.12[0.32±0.02] | 5±2[5±1] |
| YBCO | 315±15[307±10] | 4.46±0.07[0.45±0.03] | 6±2[6±2] |
| \multicolumn{4}{c}{**MgO(Sub.)/YBCO/STO(50 Å) /LSMO (YSL50) heterostructure**} |
| layer | Thickness (Å) | SLD ($10^{-5}$ Å$^{-2}$) | Roughness (Å) |
| LSMO | 275±15 [280±10] | 4.71±0.06 [0.37±0.03] | 18±5 [13±5] |
| STO | 50±4 [50±5] | 3.98±0.12 [0.33±0.02] | 10±2 [7±4] |
| YBCO | 450±25 [448±17] | 4.50±0.04 [0.45±0.03] | 15±2 [12±4] |
| \multicolumn{4}{c}{**MgO(Sub.)/YBCO/ LSMO (YL) heterostructure**} |
| layer | Thickness (Å) | SLD ($10^{-5}$ Å$^{-2}$) | Roughness (Å) |
| LSMO | 395±30 [370±20] | 4.73±0.05[0.37±0.02] | 22±3 [18±4] |
| YBCO | 309±15 [300±10] | 4.52±0.04[0.45±0.02] | 10±2 [8±2] |
| \multicolumn{4}{c}{**MgO(Sub.)/STO/ LSMO (SL) heterostructure**} |
| layer | Thickness (Å) | SLD ($10^{-5}$ Å$^{-2}$) | Roughness (Å) |
| LSMO | 230±13 [219±11] | 4.75±0.06 [0.36±0.03] | 15±6 [11±4] |
| STO | 50±7 [50±5] | 3.76±0.05 [0.32±0.02] | 4±2 [3±1] |

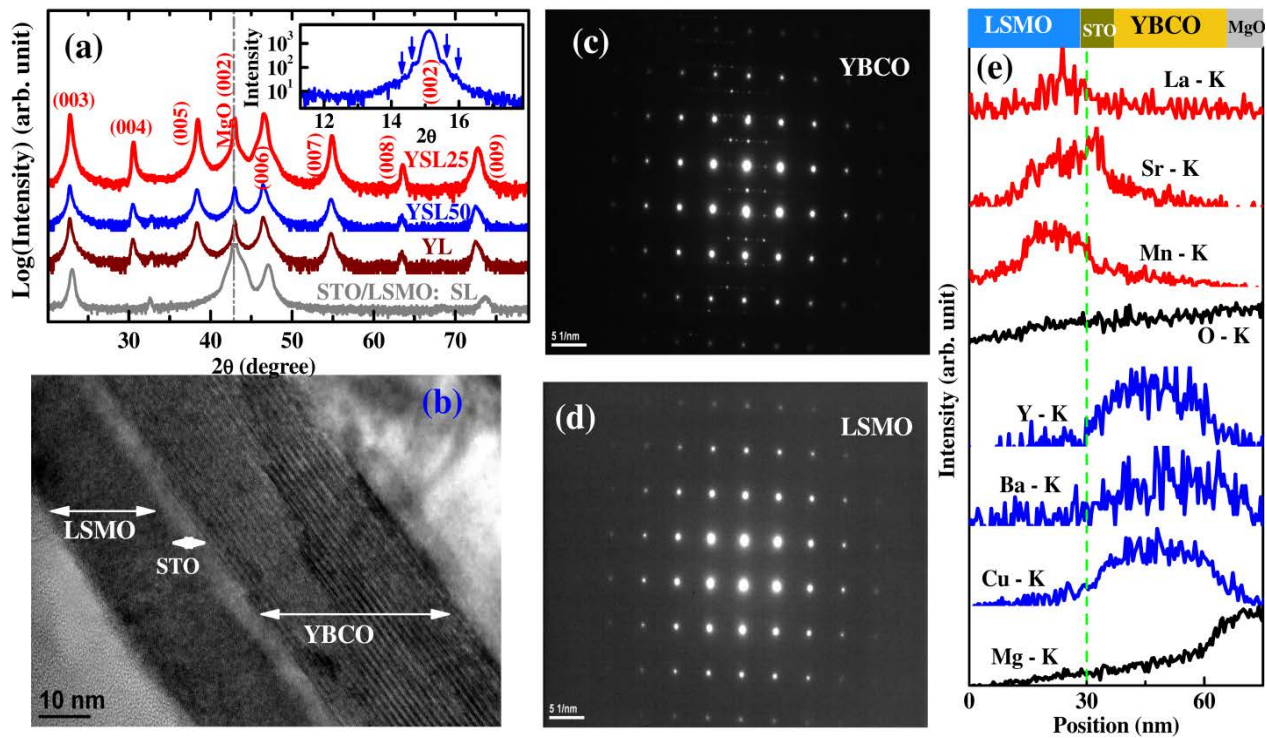

**Figure 1:** (**a**) XRD data in log scale (y-axis) from heterostructures grown on MgO. Inset shows (002) reflection of YBCO. (**b**) Cross-sectional high-resolution TEM image of a similarly grown YBCO/STO/LSMO heterostructure. Selected area electron diffraction pattern of YBCO(**c**) and LSMO (**d**). (**e**) Depth profiling of the intensity of K-edge characteristics of different elements.

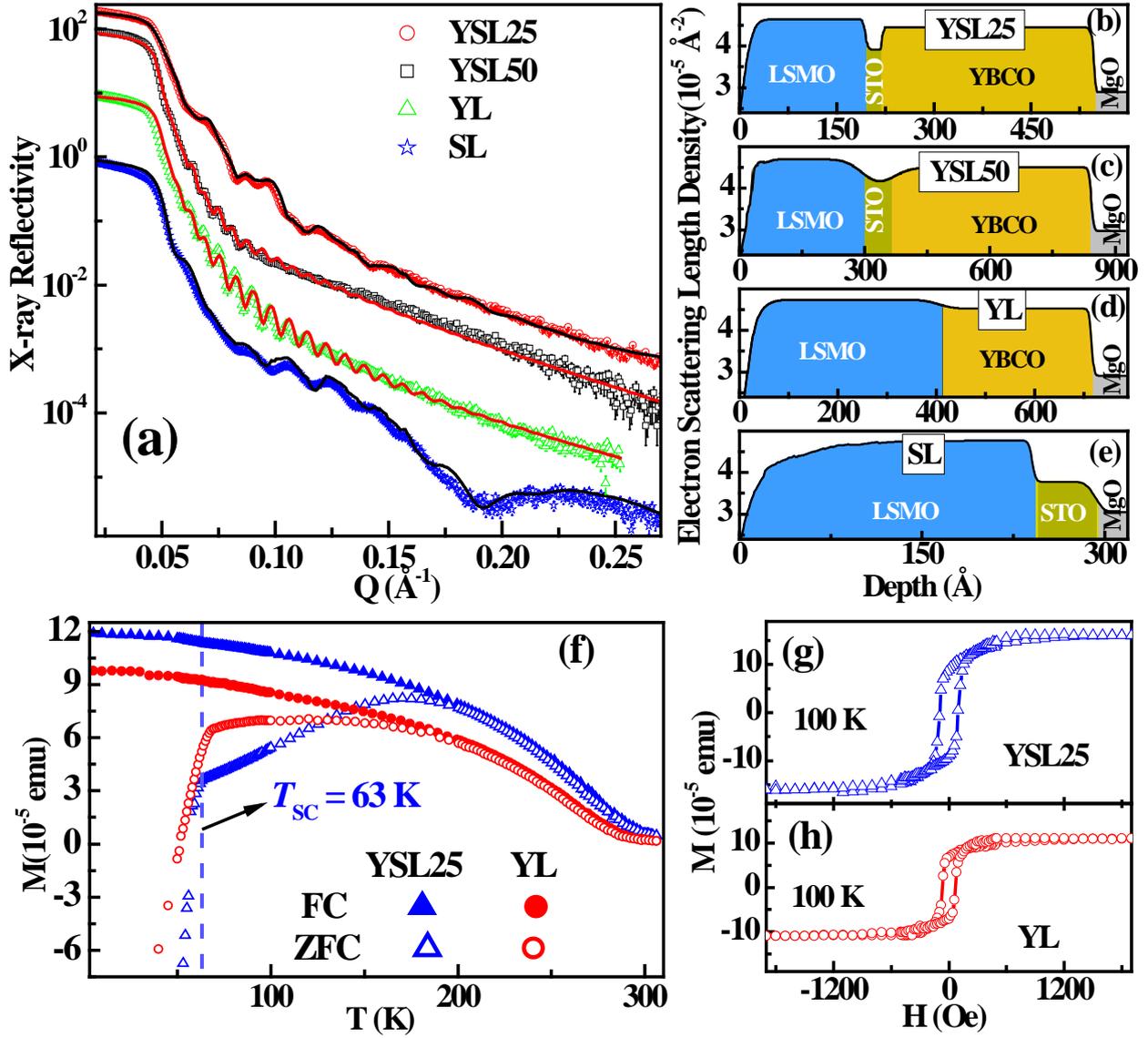

Fig. 2: (a) X-ray reflectivity (XRR) data for different (YSL25, YSL50, YL and SL) heterostructures. (b-e) The electron scattering length density (ESLD) depth profile for different heterostructures extracted from the XRR data. (f) Zero field cooled (ZFC) and field cooled (FC) magnetization of YSL25 and YSL50 in 100 Oe field. $M(H)$ curves from YSL25 (g) and YL (h) at 100 K measured by SQUID.

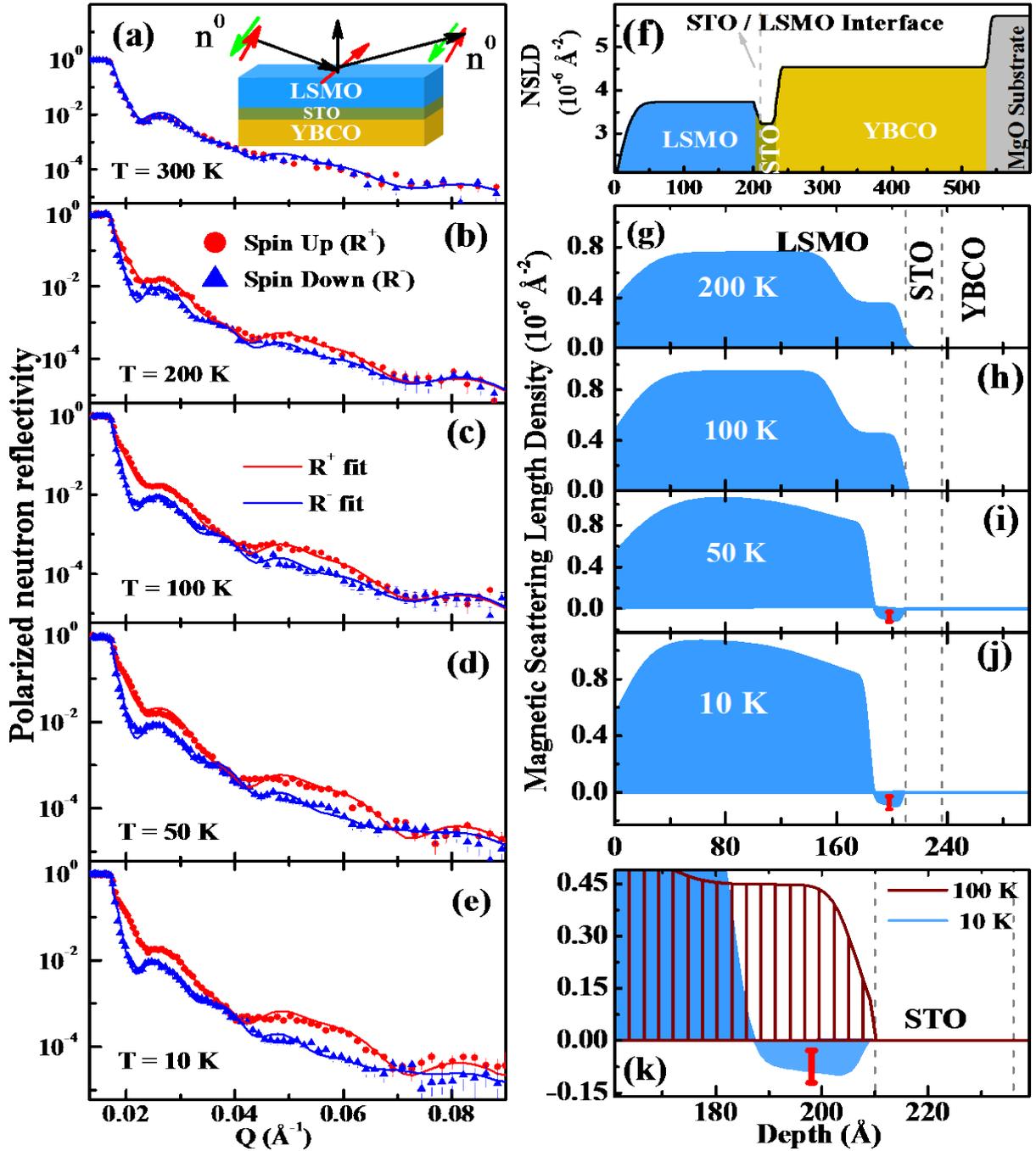

**Figure 3: (a)** PNR data from the YSL25 at 300 K. Inset of (a) shows the schematic of the PNR experiment. (**b-e**) PNR data for YSL25 at different temperatures. Inset of (d) and (e) show the ratio of $R^-(10\ K)/R^-(100\ K)$ and $R^+(10\ K)/R^+(100\ K)$ data, respectively. (**f**) Nuclear scattering length density (NSLD) depth profiles extracted from PNR data. (g-j) shows the magnetic SLD depth profile at different temperatures. (k) Comparison of MSLD near STO/LSMO interface at 100K and 10K.

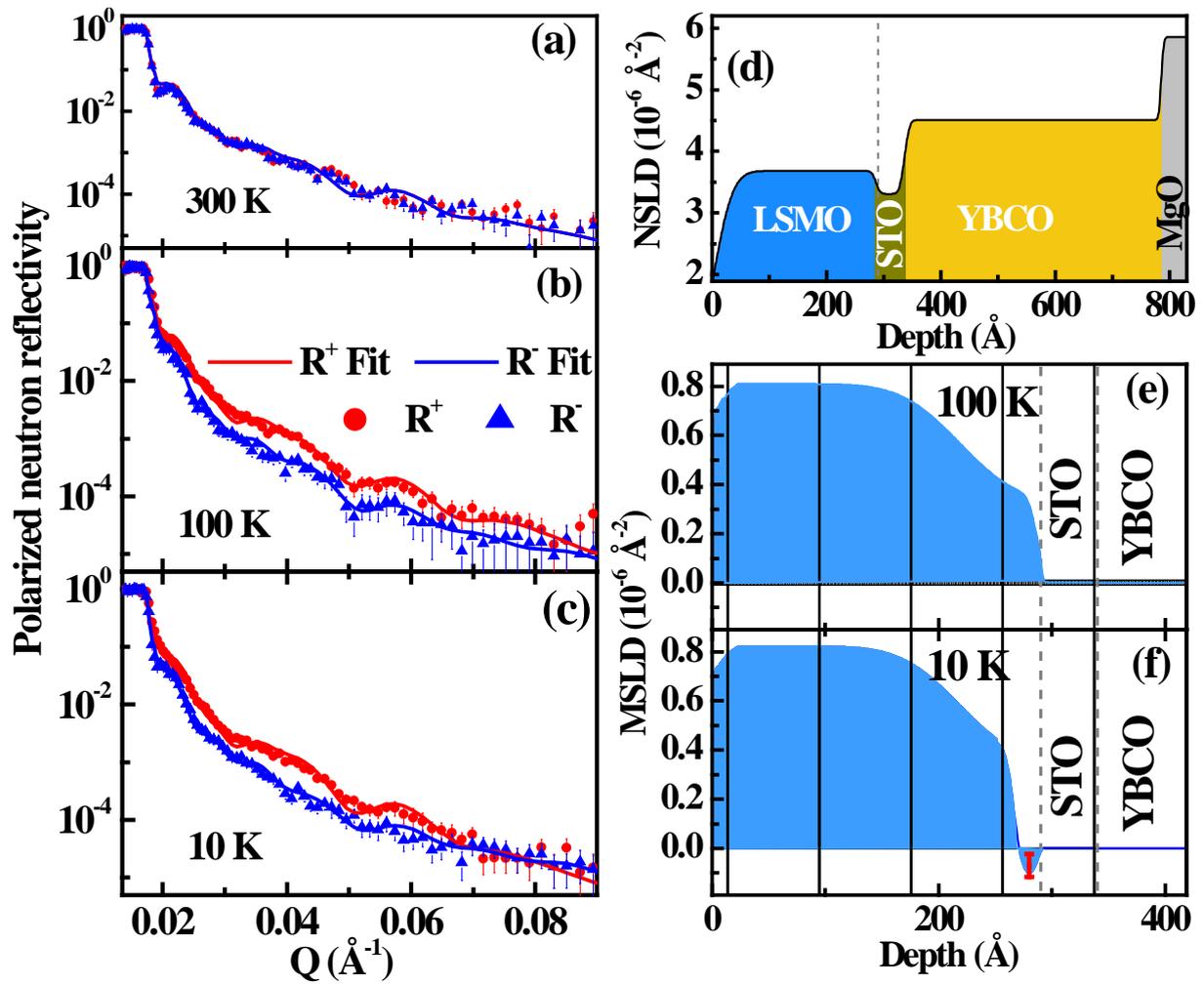

Fig. 4: (a-c) PNR data at different temperature from YSL50 heterostructure. (d) NSLD depth profile of YSL50 heterostructure. (e-f) MSLD depth profiles at 100 and 10 K for YSL50 heterostructure.

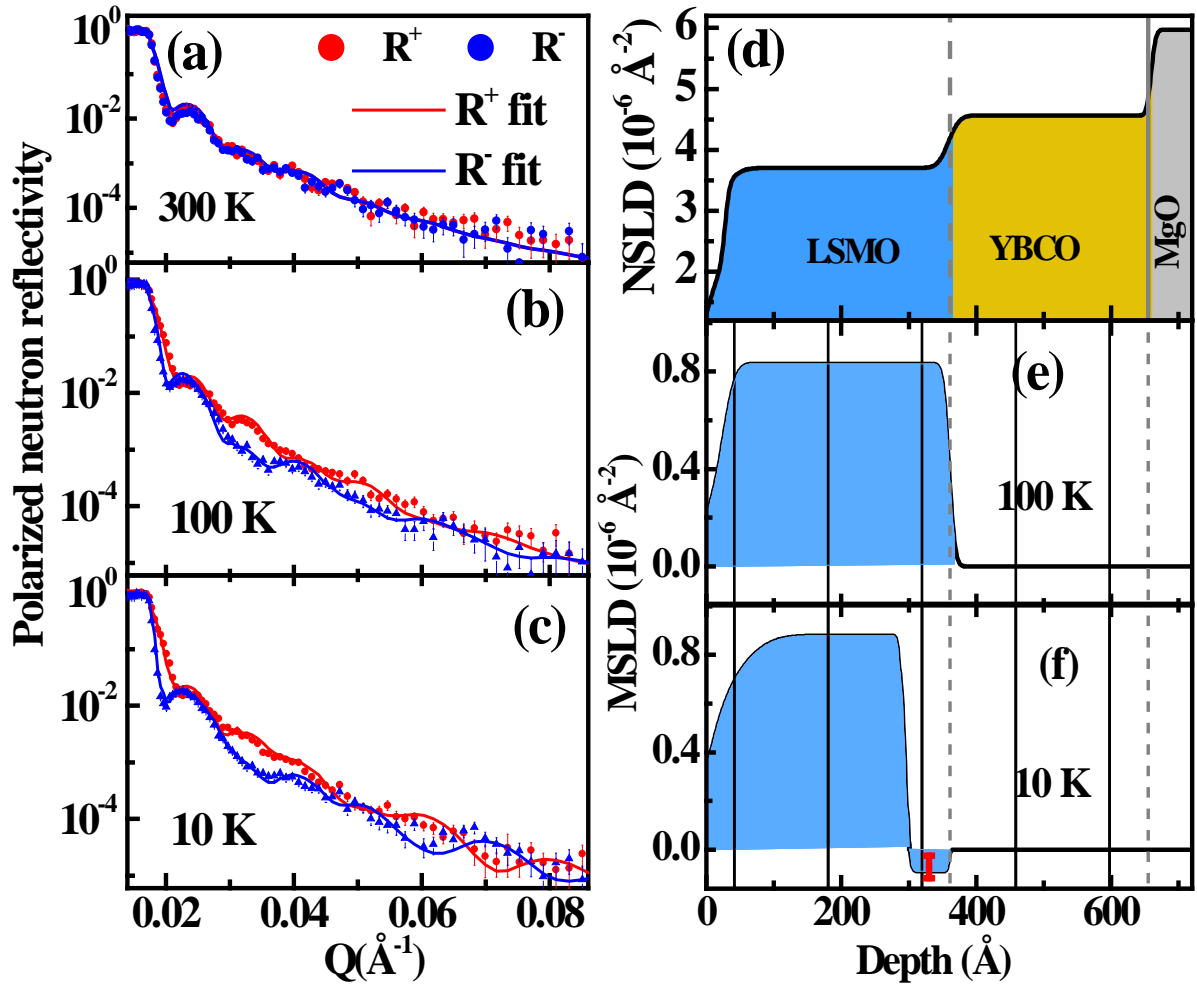

Fig. 5: (a-c) PNR data at different temperature from YL heterostructure. (d) NSLD depth profile of YL heterostructure. (e-f) MSLD depth profiles at 100 and 10 K for YL heterostructure.

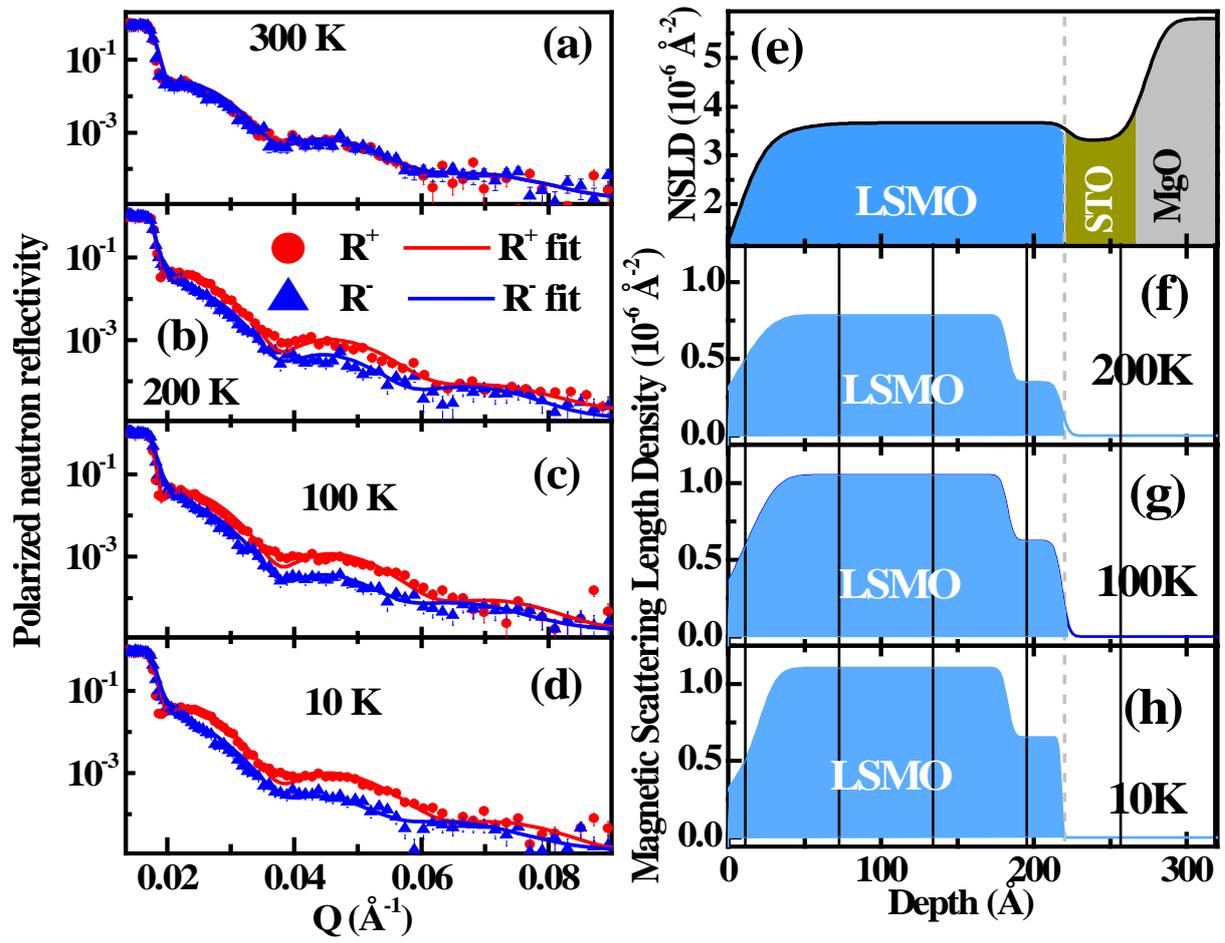

**Figure 6: (a-d)** PNR data from SL at different temperatures. (**e**) NSLD depth profile of SL heterostructure. (f-h) MSLD depth profile extracted from the PNR data at different temperatures.

# SUPPLEMENTARY INFORMATION

# Superconductivity-driven magnetization modulation in YBa$_2$Cu$_3$O$_{7-\delta}$/SrTiO$_3$/La$_{0.67}$Sr$_{0.33}$MnO$_3$ heterostructures

Surendra Singh[1,2,*], Harsh Bhatt[1], Yogesh Kumar[1], C. L. Prajapat[3], B. Satpati[4], C. J. Kinane[5], S. Langridge[5], G. Ravikumar[2,6] and S. Basu[1,2]

[1]Solid State Physics Division, Bhabha Atomic Research Centre, Mumbai 400085 India
[2]Homi Bhabha National Institute, Anushaktinager, Mumbai 400094 India
[3]Technical Physics Division, Bhabha Atomic Research Centre, Mumbai 400085 India
[4]Surface Physics and Material Science Division, Saha Institute of Nuclear Physics, 1/AF, Bidhannagar, Kolkata 700064, India
[5]ISIS-Neutron and Muon Source, Rutherford Appleton Laboratory, Didcot, Oxon OX11 0QX, United Kingdom
[6] Scientific Information Resource Division, Bhabha Atomic Research Centre, Mumbai 400085 India
*surendra@barc.gov.in

## 1. Sample Growth:

Pulsed laser deposition (PLD) was used to grow different bilayers and trilayers of La$_{0.67}$Sr$_{0.33}$MnO$_3$ (LSMO), SrTiO$_3$ (STO) and YBa$_2$Cu$_3$O$_{7-\delta}$ (YBCO) on MgO (001) single crystal substrates. Before growing actual hybrid YBCO/LSMO and YBCO/STO/LSMO heterostructures, we have grown single layer of each oxide on MgO (001) to study the effect of strain on the growth of these heterostructures. We have discussed four hybrid heterostructures and single oxide layers with following structures (Table S1).

A base pressure of ~10$^{-6}$ mbar was achieved in the PLD chamber before every deposition. An excimer laser (KrF, λ=248 nm, pulse width = 20 nm) was used to ablate high density targets of LSMO, YBCO and STO. The laser fluence was fixed at 3 J/cm$^2$ and an optimized pulse repetition rate of 2 Hz was used for the deposition under an oxygen partial pressure of 0.5 mbar. The film deposition was carried out at an optimized substrate temperature of 800 °C and was followed by post deposition annealing for 60 minutes at 550 °C in 1000 mbar O$_2$.

Table S1: Structure of different oxide heterostructures grown on MgO single crystal substrate with their ferromagnetic (FM) and superconducting (SC) order temperatures.

| Sample | Sample code | FM transition temperature, $T_C$ (K) | SC transition temperature, $T_{SC}$ (K) |
|---|---|---|---|
| MgO(substrate)/YBCO/STO(~25 Å)/LSMO | YSL25 | 290 | 63 |
| MgO(substrate)/YBCO/STO(~50 Å)/LSMO | YSL50 | 290 | 63 |
| MgO(substrate)/YBCO/LSMO | YL | 290 | 63 |
| MgO(substrate)/ STO/LSMO | SL | 290 | - |
| MgO(substrate)/ YBCO | - | - | 88 |
| MgO(substrate)/LSMO | - | 310 | - |
| MgO(substrate)/ STO | - | - | - |

## 2. X-ray Diffraction

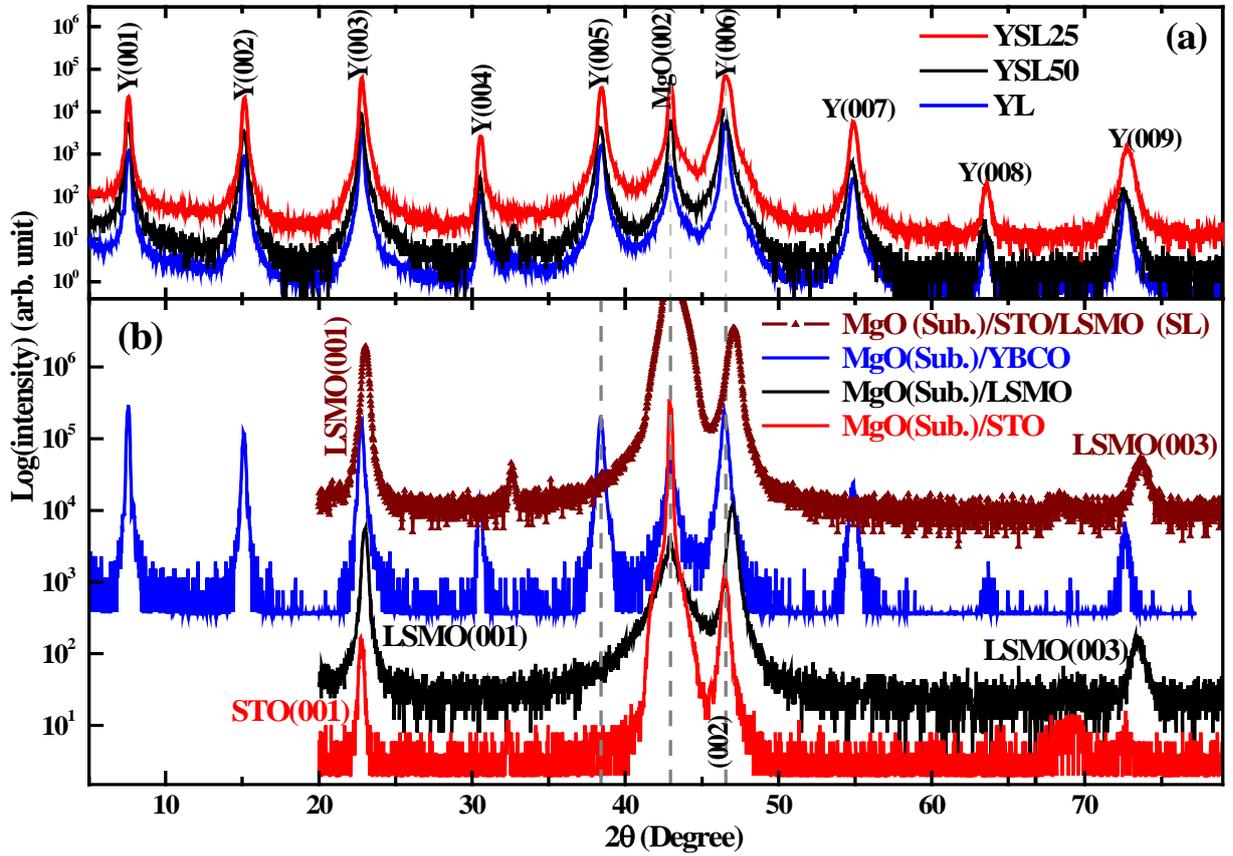

Fig. S1: (a) X-ray diffraction data from MgO/YBCO/STO (25 Å)/LSMO (YSL25), MgO/YBCO/STO (50 Å)/LSMO (YSL50), MgO/YBCO/LSMO (YL) systems in log scale. (b) X-ray diffraction data in log scale from MgO/LSMO, MgO/STO, MgO/YBCO single layers and MgO/STO/LSMO (SL) bilayer. XRD profile from different films are shifted vertically for better visualization.

The high degree of crystallinity of the heterostructures was ensured by the X-ray diffraction (XRD) measurements. Before deposition of heterostructures we optimized the growth of the LSMO, STO and YBCO layers individually (single layer) on MgO (001) substrate. XRD patterns from the two trilayers (YSL25, YSL50) and the bilayer (YL) are shown in Fig. S1(a). Fig. S1(b) shows the XRD patterns of the MgO/LSMO, MgO/STO, MgO/YBCO films and SL bilayer. The XRD data clearly suggests highly preferential growth in (00l) direction. We have used these Bragg reflections to estimate the strain for different layers grown on MgO (001) substrate and on other oxide layers for bilayer and trilayer heterostructures. The estimated strain values are given in Table S2.

**Table S2**: Estimated strain values for different oxide[1-3] layers grown as single layer, bilayer and trilayer on MgO substrate.

|      | MgO/ Single Layer | MgO/ STO/LSMO (SL) | MgO/ YBCO/LSMO (YL) | MgO/ YBCO /STO(25Å)/ LSMO (YSL25) | MgO/ YBCO /STO(50Å)/ LSMO (YSL50) |
|------|-------------------|--------------------|---------------------|-----------------------------------|-----------------------------------|
| LSMO | -0.39%            | -0.41%             | 0.39%               | 0.52%                             | 0.03%                             |
| STO  | -0.08%            | -1.54%             | -                   | -0.15%                            | -0.64%                            |
| YBCO | -0.12%            | -                  | -0.15%              | -0.07%                            | -0.26%                            |

## 3. XRR measurements

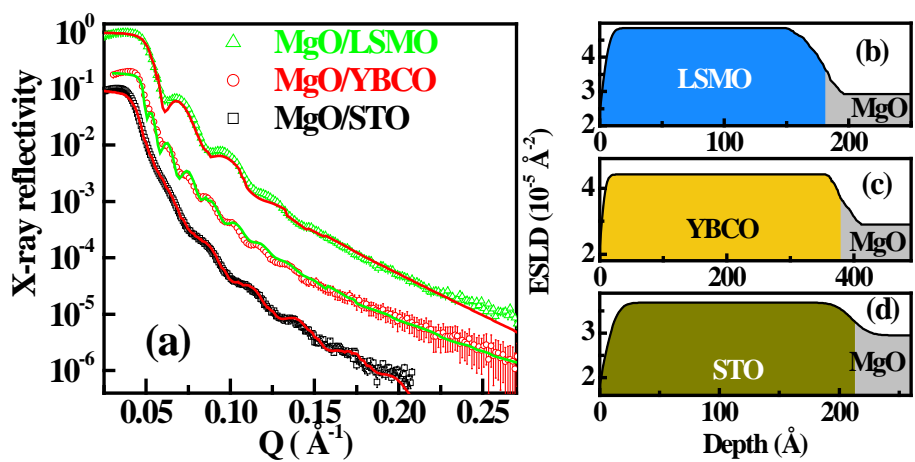

Fig. S2: (**a**) XRR data from single layer (LSMO, STO and YBCO) grown on MgO substrates. XRR profile from different films are shifted vertically for better visualization. (**b**) Electron scattering length density (ESLD) depth profiles for corresponding single layers, which best fitted XRR data.

Fig. S2(a) shows the XRR data from individual single layers of STO, YBCO and LSMO grown on MgO substrate. Reflectivity for bilayer and trilayer are shifted for better visualization. Fig. S2(b-d) show the corresponding electron scattering length density (ESLD) depth profiles which best fitted XRR data shown in Fig. S2(a).

## 4. PNR Measurements

In PNR, magnetic depth profiles are directly related to the difference between spin dependent reflectivity ($R^+ - R^-$) and hence a quantity called normalized spin asymmetry (NSA) defined as NSA $= \frac{(R^+ - R^-)}{(R^+ + R^-)}$, is calculated and compared for different magnetic depth profiles. Different magnetic models considered in this paper are statistically compared for quality of each fits using[4]: $\chi^2 = \sum_i [NSA_{exp}(Q_i) - NSA_{the}(Q_i)]^2$, where $NSA_{exp}(Q_i)$ and $NSA_{the}(Q_i)$ are the NSA data points and NSA for fitted model at momentum transfer $Q_i$, respectively. The smaller is the $\chi^2$, the better the fit. Thus, the $\chi^2$ values corresponding to each MSLD model, where we have compared different models, is given in different reflectivity plots.

*PNR from YSL25 heterostructure:*

Given the mixed reports for magnetization modulation for STO/LSMO (LSMO on STO) interface[5-7] we have considered several magnetization depth profile models for YBCO/STO/LSMO heterostructure (YSL25) to fit PNR data at 100 K. The models considered are (i) uniform magnetization in the entire LSMO layer (ii) uniform magnetization with a magnetic dead layer at STO/LSMO interface (iii) an enhanced magnetization at STO/LSMO interface, and (iv) a reduced magnetization at STO/LSMO interface.

A detailed comparison of the fits to PNR data at 100 K from YSL25 heterostructure with different magnetization depth profile models are shown in Fig. S3. Solid lines in Fig. S3 (a)-(d) show the fit to PNR data at 100 K, considering the above mentioned magnetization depth profiles shown in Fig. S3 (f)-(i). Fig. S3 (e) shows NSLD profile obtained from the fit of PNR data at 300 K for YSL25 heterostructure, which was kept fixed for analyzing PNR data at low temperatures. We have varied the thickness of the interface layer considering total thickness of the layer constant and the magnetization profiles shown for different models corresponds to the best fit (lowest $\chi^2$) obtained for the particular model. The magnetization is other parameters in different model which was varied to get best fit for different models.

We obtained better fit (lowest $\chi^2$) for magnetization model with a reduced magnetization at STO/LSMO interface (Fig. S3 (d) and (i)). For other models we observed increase in $\chi^2$ in comparison to the fit obtained assuming reduced magnetization at interface. $\chi^2$ for different magnetization models discussed above are shown in Fig. S3. Thus it is evident from Fig. S3 that reduced magnetization for LSMO layer at STO/LSMO interface best fitted PNR data at 100 K for YSL25 heterostructure.

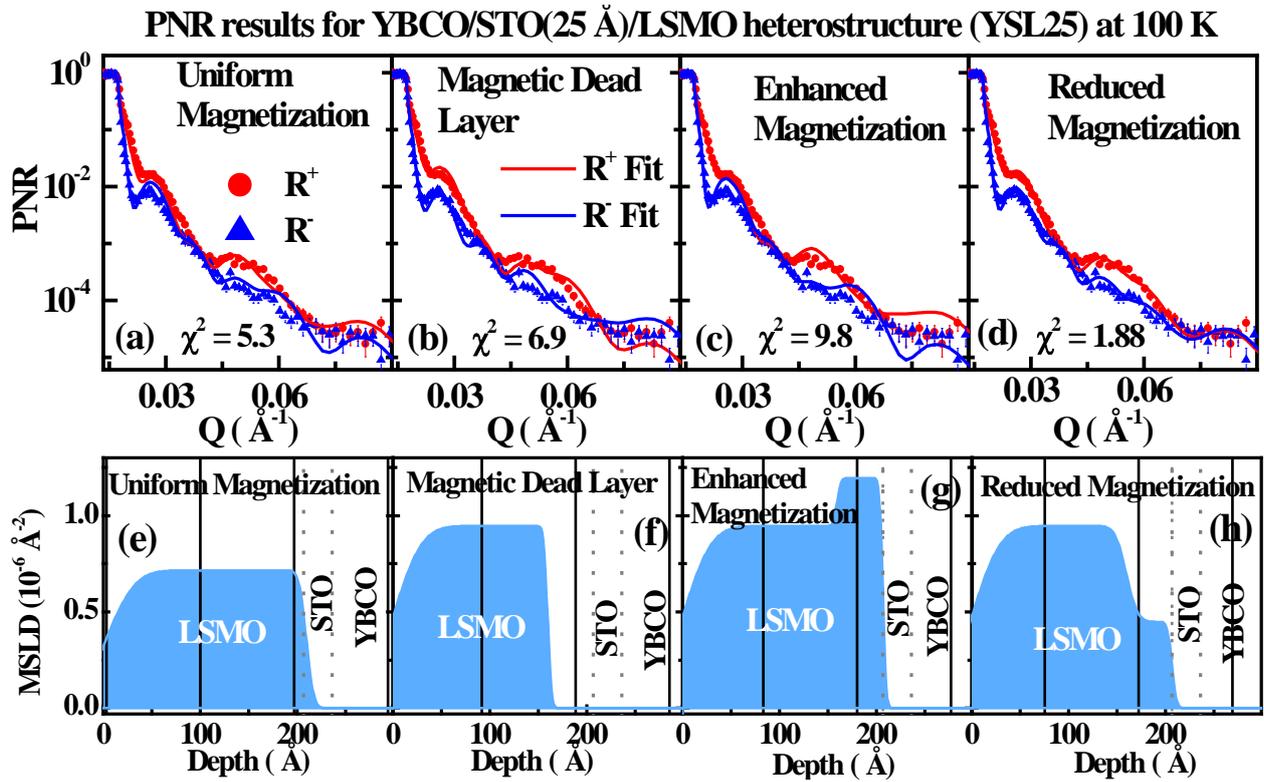

Fig. S3: (a)-(d) Polarized neutron reflectivity (PNR) data (spin up, $R^+$ and spin down, $R^-$) from the YBCO/STO/LSMO (YSL25) heterostructure at 100 K. (e)-(h) Different magnetization depth profile models used to fit (solid lines) PNR data in (a) to (d).

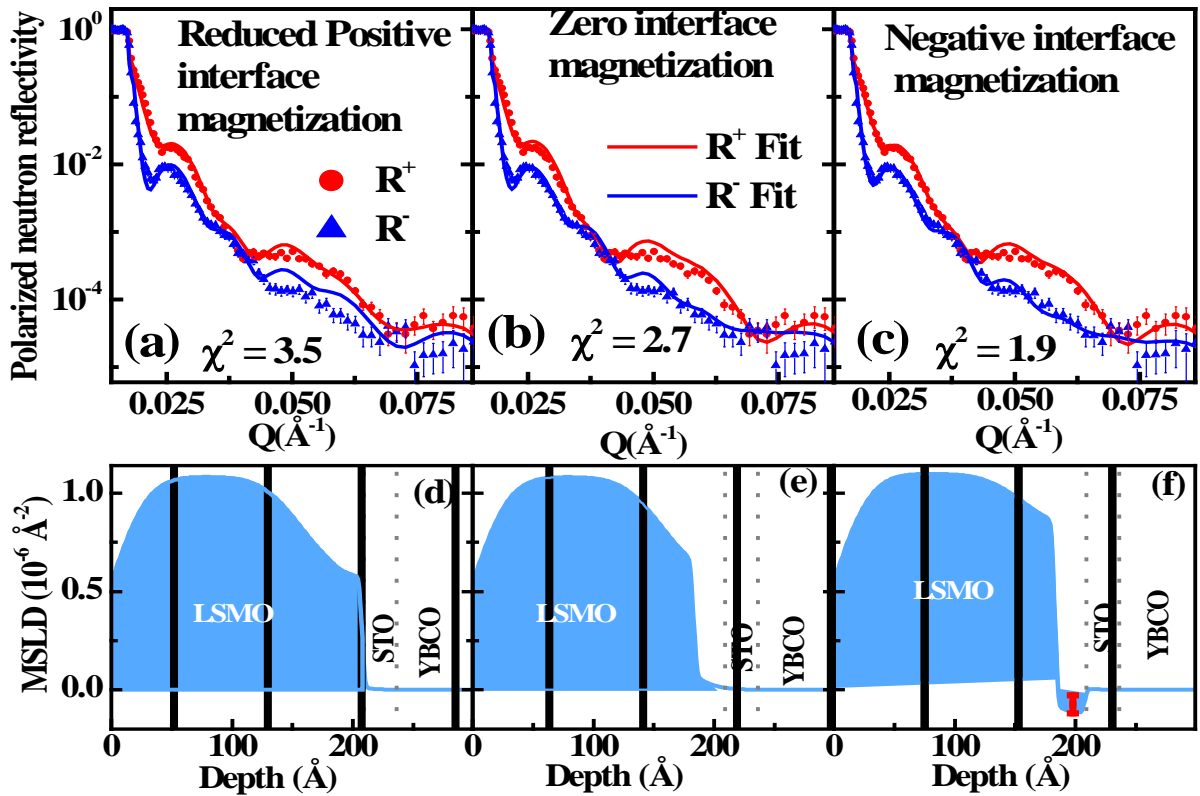

Fig. S4: (a-c) Comparison of PNR data at 10 K from YSL25 heterostructure and fitting assuming different magnetization models e.g. assuming positive (d), zero (e) and negative magnetization (f) for LSMO at the interface with STO.

Comparison of magnetization models e.g. positive, zero and negative magnetization of interface LSMO layer of thickness ~ 30 Å with fit to PNR data at 10 K from YSL25 heterostructure are shown in Fig S4 (a), (b) and (c). We found lowest $\chi^2$ (~1.9) as compared to that of positive ($\chi^2$ ~3.5) and zero ($\chi^2$ ~2.7) magnetization models. Thus larger $\chi^2$ value for the magnetization model with positive and zero magnetization at STO/LSMO interface (Fig. S4(d) and (e)) suggesting that negative magnetization model (Fig. S4 (f)) at STO/LSMO interface in YSL heterostructure better fitted the PNR data at 10 K.

A comparison of reflectivity data as a ratio of $R^+(10K)/R^+(100K)$ and $R^-(10K)/R^-(100K)$ are shown in Fig. S5 (a) and (b) respectively. Fig. S5(c) shows the comparison of MSDL depth profiles at 100 and 10 K which best fitted the data.

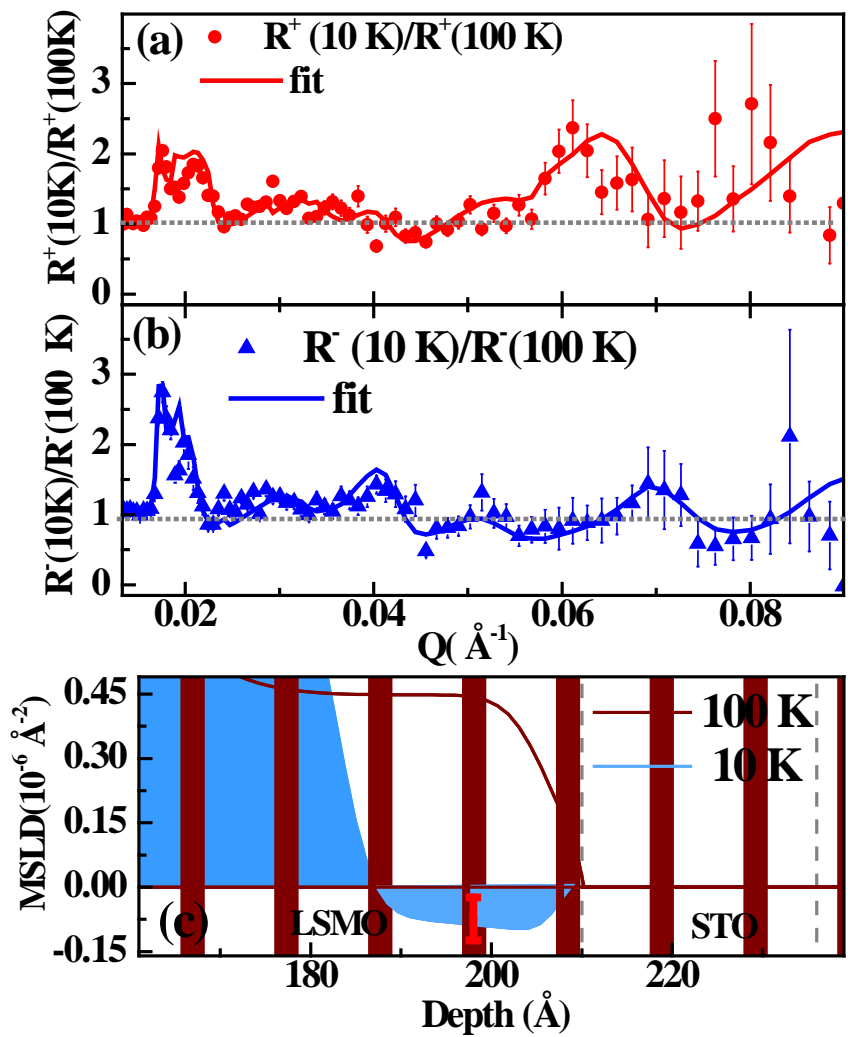

Fig. S5: Reflectivity ratio of (a) $R^+(10\ K)/R^+(100\ K)$ and (b) $R^-(10\ K)/R^-(100\ K)$ data (symbols) and corresponding fit (solid line). (c) Comparison of MSLD depth profiles at 100 and 10K for YSL25 near STO/LSMO interface.

*PNR from STO/LSMO (SL) heterostructure:*

Reduced magnetization at STO/LSMO interface was further verified by measuring PNR data on similarly grown STO/LSMO bilayer (SL) on MgO substrate. Fig. S6 shows the PNR results from SL at 100 K (Fig. S6 (a-c)) and 10 K (Fig. S6 (d-f)), considering different MSLD depth profiles. It is clear from Fig. S6 that reduced magnetization for LSMO layer near STO/LSMO interface best fitted the PNR data. Statistical variation of $\chi^2$ for different magnetization models are given in the Fig. S6.

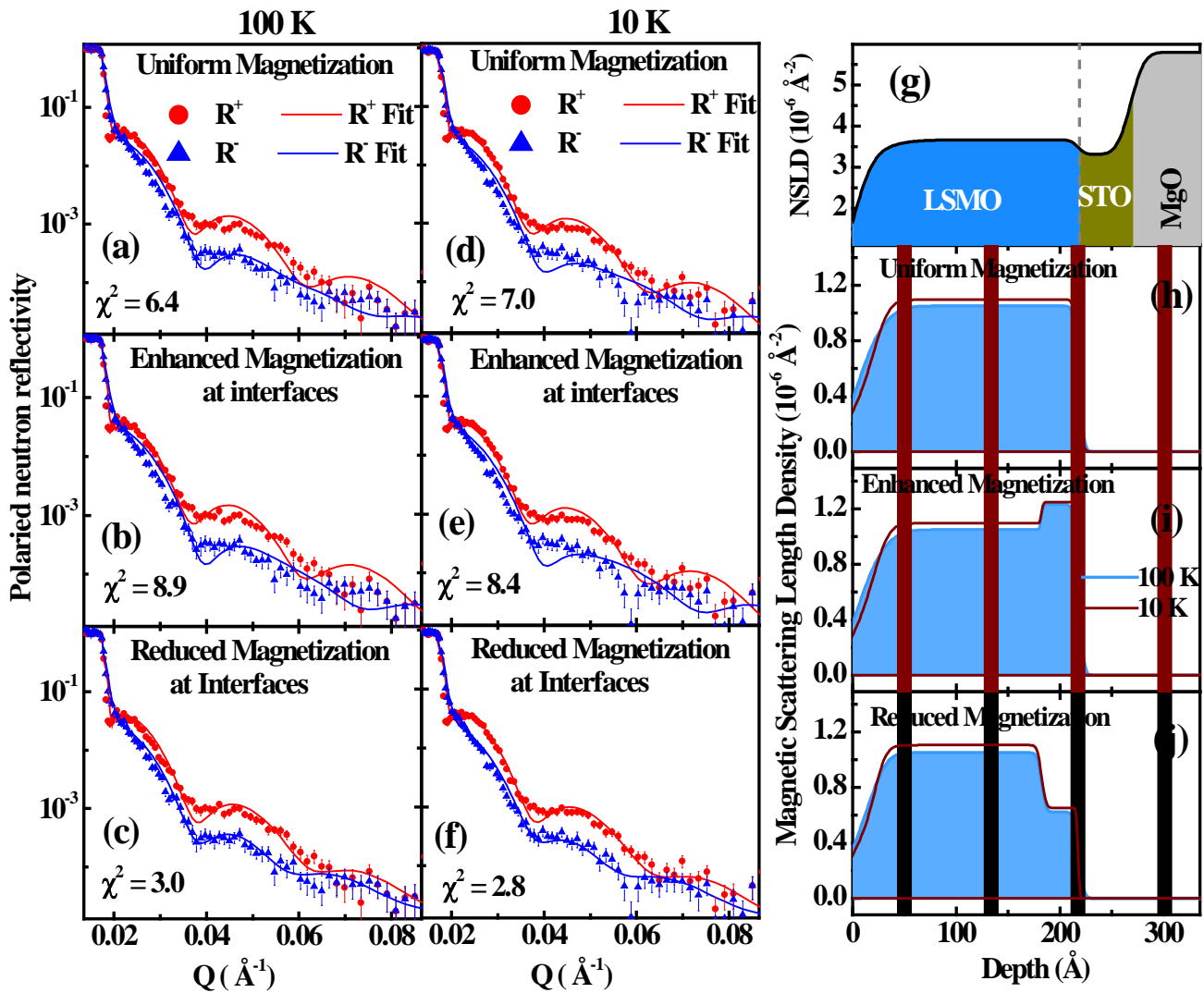

Fig. S6: PNR data (spin up, $R^+$ and spin down, $R^-$) from the STO/LSMO (SL) heterostructure at 100 K (a-c) and at 10 K (d-f). (g) Nuclear scattering length density (NSLD) depth profile of the sample extracted from PNR data at 300 K for SL, suggesting well defined heterostructure. (h)-(j) Different magnetization depth profile models used to fit (solid lines) PNR data in (a) to (f).